# An Open-Source Simulink–Based Program for Simulating Power Systems Integrated with Renewable Energy Sources

Ismael Abdulrahman, ikabdulrah42@students.tntech.edu
Tennessee Technological University, Department of Electrical and Computer Engineering, Cookeville, TN, USA
Information System Engineering Department, Erbil Technical Engineering College, Erbil Polytechnic University, Iraq

*Note*: The author would like to share the program publicly (open-source). The link to the files can be obtained from [17]. A short video was created while running the program and is available in [18].

*Abstract*—This paper presents an open-source Simulink-based program developed for simulating power systems integrated with renewable energy sources (RES). The generic model of a photovoltaic, wind turbine, and battery energy storage is used for the RES. The program can be used for educational and research studies. It comes with several important subjects in power systems including power system modeling and integration, linearization, modal analysis, participation factor analysis, controller selection using residue analysis, and frequency response analysis. IEEE 68–bus dynamic test system is used to verify the performance of the program. The results show the efficiency and speed of the program to simulate a large-scale modern power system.

*Index Terms*— Renewable Energy Sources (RES), Power System Modeling, Power System Integration, Simulink.

## 1. Introduction

With the growing penetration level of renewable energy sources (RESs) such as wind turbines and solar photovoltaic (PV) plants, there is an increasing need by researchers to develop programs that include modeling of new power-electronic-based sources. Since mid-2014, major commercial and professional software platforms have added the generic model of RESs to their libraries based on the plant model developed primarily by Western Electricity Coordinating Council (WECC) and extended later by Electric Power Research Institute (EPRI) for stability studies [1-5]. Some of these platforms are GE PSLF, PSS\E, PowerWorld Simulator, etc. These programs are professional, well-coded, computationally efficient, fast, and accurate. However, commercial software has four main drawbacks [6]–[8]: (1) its components are "closed-source"; the user is not allowed to edit the code sources except for some special cases (2) it is expensive (3) it provides less educational due to the closed-source nature of its components (4) it requires training courses. On the other hand, MATLAB\Simulink is a high-level language platform, less expensive, available at almost all universities, and students and researchers are familiar with it. Although Simulink has added block elements in its library for the wind turbine, solar plant, and battery systems, however, none of these components are based on the generic model developed recently by EPRI and WECC. Besides, these elements are based on single-input single-output (SISO) which requires a huge amount of computation time and effort for simulating a multi-machine power system.

The study in [9] presented a Simulink-based program for integrating a generic model of PV plant into a two-area power system. The developed program, however, takes a quite long time to simulate the system owing to the existence of an algebraic loop problem in its model. In other words, the program in [9] can only be used for studying the dynamic behavior of small-scale power systems, and there is a need to develop a program that can be employed for simulating large-scale power systems integrated with renewable energy resources. Reference [10] introduces a Simulink-based program for simulating large-scale power systems. This program is developed to simulate a conventional power system by considering all generators as synchronous machines. Modern power systems are hybrid systems that contain both synchronous machines and renewable energy plants. PSAT toolbox [6] is another open-source MATLAB-based package used for integrating renewable energy sources. However, modeling of renewable energy sources using the recent EPRI and WECC presentation is not considered in this toolbox. Each component represents only one RES source and the signals connected to these sources represent only one input or output data (scalar-based signals).

This paper presents a vectorized approach to model, simulate, and linearize modern power systems equipped with green-energy generators using Simulink. Modal analysis, participation factor analysis and visualization, frequency response analysis, and optimal location of the controller are other subjects considered in this paper using the program. The renewable energy sources are modeled using the generic model adopted by WECC and EPRI. To speed up the simulation process, all the algebraic equations are absorbed into the admittance matrix of the system. The signals connecting the RES sources and other parts of the system can be either scalars or vectors.

The rest of this paper is presented as follows. In Section 2, the system modeling is described including various parts of the generic model of renewable energy sources, synchronous generator, network Simulink model, and initial condition calculation. Next, a simulation example is presented in Section 3 which includes several subsections such as time-domain simulation, modal analysis, participation factor analysis, frequency response analysis, and control design. Finally, the conclusion is introduced in Section 4.

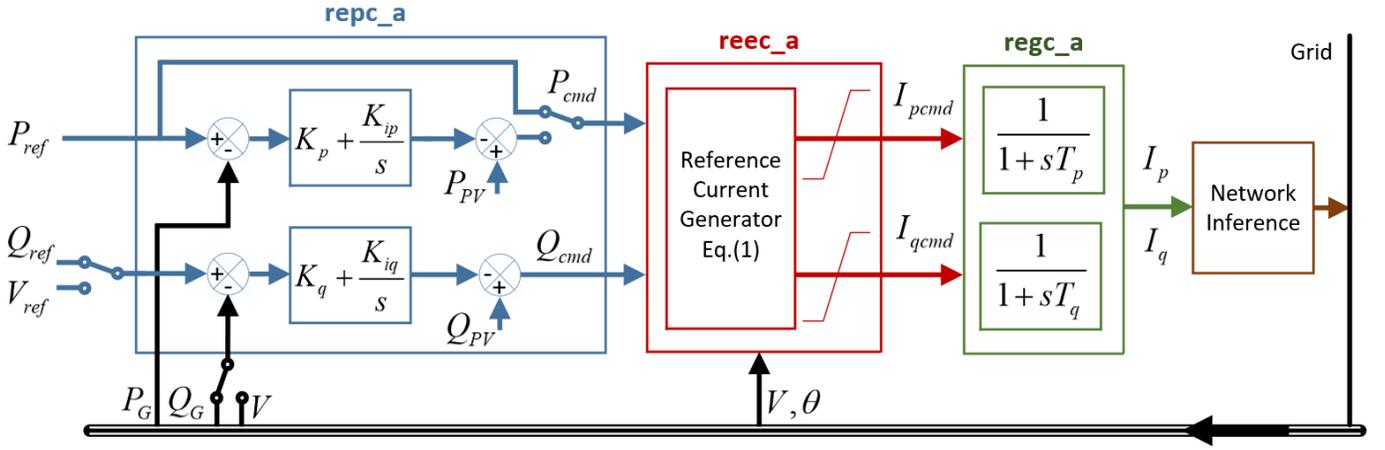

Fig. 1. The generic model of RES [1-3], [8]

## 2. SYSTEM MODELING

### 2.1 Generic Model of RES

The first generic renewable energy systems models were developed between 2010 to 2013, whereas the second-generation models were developed by EPRI in 2015 [1]. These models can be used to represent Type 1–4 WTG wind turbines, Photovoltaic (PV) solar plants, and Battery Energy Storage Systems (BESS). The generic model of RESs is based on Type 4 WTG wind turbines shown in Fig. 1. and can be used for modeling wind turbines, PV plants, and BESSs for studies that their main concern is the dynamic behavior of the system. The structure of this model consists of the following parts [1], [11]:

a) regc_a: this part of the model represents the convertor of renewable energy plants. It has two inputs namely $I_{pcmd}$ and $I_{qcmd}$ denoting the active and reactive command currents, respectively, and two outputs namely $I_p$ and $I_q$ referring to the current injections into the grid. Mathematically, two identical low-pass filters with a fast time constant (usually in the range of 0.01 to 0.02s) are used to model the convertor.

b) reec_a: this part of the model represents the electrical control of the renewable energy plant. It has two inputs namely $P_{cmd}$, and $Q_{cmd}$, denoting the command active and reactive powers, and two outputs namely $I_{pcmd}$ and $I_{qcmd}$, respectively. The command currents can be computed using the following equation:

$$\begin{bmatrix} I_{pcom} \\ I_{qcom} \end{bmatrix} = \begin{bmatrix} V_d & V_q \\ V_q & -V_d \end{bmatrix}^{-1} \begin{bmatrix} P_{cmd} \\ Q_{cmd} \end{bmatrix} \quad (1)$$

where $V_d$ and $V_q$ are the $dq$-components of the RES terminal voltage ($V$), $I_{dcmd}$ and $I_{qcmd}$ are the $dq$-components of the generated command currents used as signal inputs to the converter, $P_{cmd}$ and $Q_{cmd}$ are the command power signals, which represent the generated RES active and reactive powers.

c) repc_a: this part of the model represents the plant controller. The upper path uses active power reference as an input to the block. Nominal frequency can also be used in this loop as a reference signal [1]. The lower path of the plant uses either reactive power reference or voltage reference as an input to the block. There is feedback from the grid to the plant which can be either the grid active\reactive power or the terminal voltage, depending on the loop control and the main objective of this control [8]. Mathematically, one or two PI-controllers are used to generate active and reactive power commands. Reference [11] uses the reactive power loop control with a PI-controller, whereas [1-3] present a two-loop control scheme. An additional damping controller can be added to one of the loops to enhance the stability of the system [9]. The reactive power loop can be controlled to maintain power factor constant or to control the terminal voltage of the plant. For more details refer to [12].

Note that the model names used in the commercial software platforms are slightly different owing to the naming conversion utilized by these software vendors. Table I shows the difference between these model names used by the most widely used tools in WECC and US regions [1]. This paper employs the same names used by GE PSLF.

TABLE I model names used in the commercial platforms [1]

| GE PSLF | MODEL NAME | |
|---|---|---|
| | Siemens PSSE | PowerWorld Simulator |
| regc_a | REGCAU1 (V33); REGCA1 (V34) | REGC_A |
| reec_a | REECAU1 (V33); REECA1 (V34) | REEC_A |
| repc_a | REPCTAU1 & REPCAU1 (V33); REPCTA1 & REPCA1 (V34) | REPC_A |

### 2.2 Synchronous Generator, Turbines, and Exciters

The synchronous generators are modeled using the detailed sixth-order representation equipped with exciters and turbines. The abbreviations and mathematical description of this part of the system can be obtained in Appendix I and II, respectively [13].

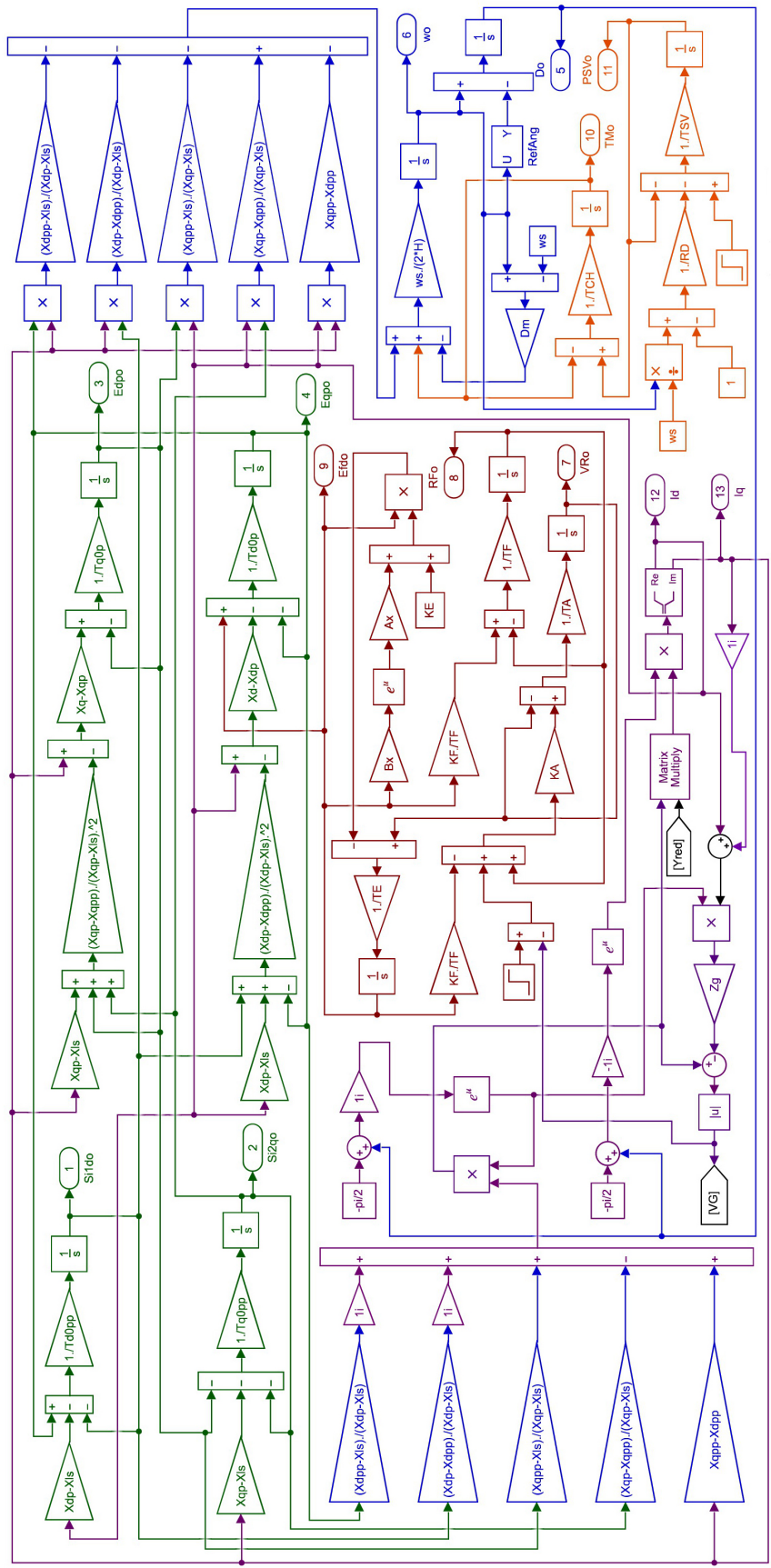

Fig. 2a. The developed Simulink program to integrate the generic model of RES – Part 1

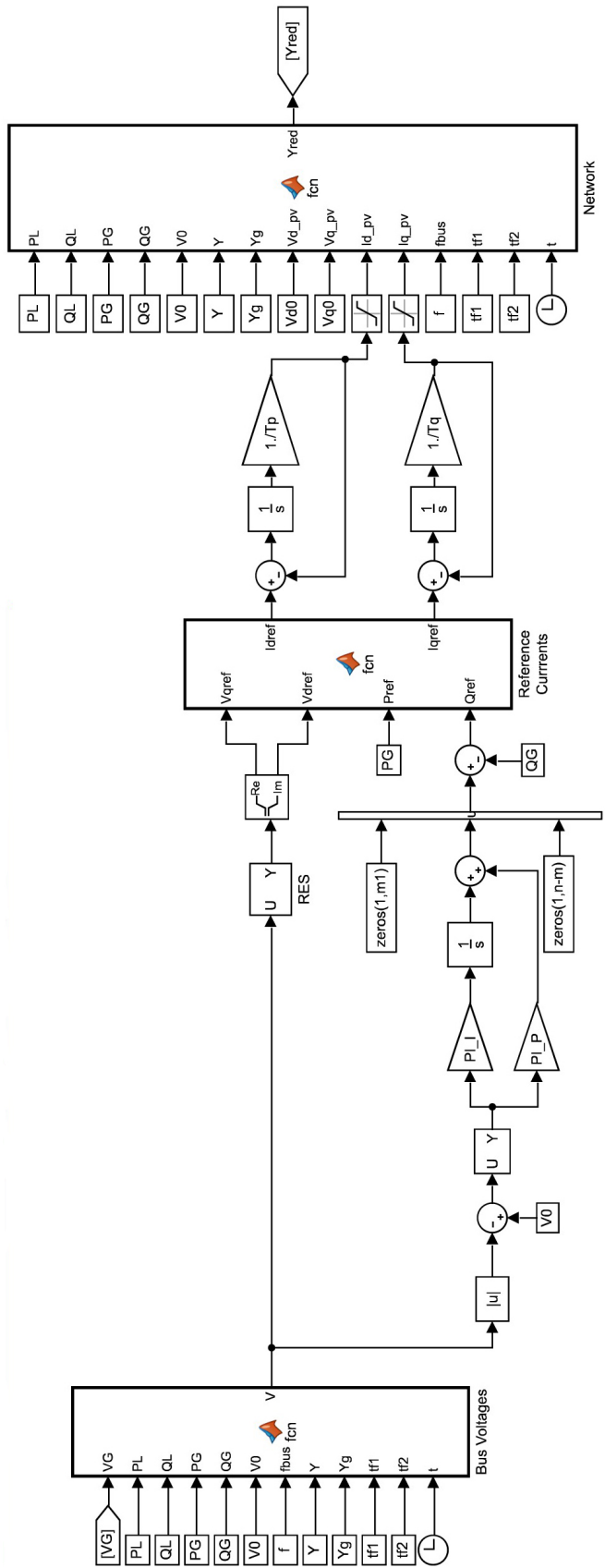

Fig. 2b. The developed Simulink program to integrate the generic model of RES – Part 2

## 2.3 Simulink Model

To avoid algebraic loops in the Simulink model and to speed up the simulation, the loads are treated as constant impedances and absorbed into the admittance matrix. The synchronous generators are represented by the differential equations (10)-(15) in Appendix II whereas the renewable energy plants are modeled using the generic model of RES. The RES model adopts one loop control scheme represented by the reactive control section described in [11, p. 220]. The block diagram of the Simulink model is shown in Fig. 2a-b. Each differential equation is highlighted by giving a specific color. The three MATLAB function blocks in Fig. 2a-b are described briefly as follows.

The first MATLAB–function block on the left of Fig. 2b namely "Voltage Bus" is aimed to calculate the voltage magnitudes for the buses that are not connected to the synchronous generators (load buses and RES buses) using (4) as follows [10]:

$$\begin{bmatrix} I_G \\ 0 \end{bmatrix} = \begin{bmatrix} Y_{11} & Y_{12} \\ Y_{21} & Y_{22} \end{bmatrix} \begin{bmatrix} V_G \\ V_L \end{bmatrix} \quad (2)$$

$$\begin{cases} I_G = Y_{11}V_G + Y_{12}V_L \\ 0 = Y_{21}V_G + Y_{22}V_L \end{cases} \quad (3)$$

From the second equation in (3), we can find $V_L$:

$$V_L = -Y_{22}^{-1} Y_{21} V_G \quad (4)$$

where $Y_{11}$, $Y_{12}$, $Y_{21}$, and $Y_{22}$ refer to submatrices of the augmented admittance matrix with dimensions $m*m$, $m*(n-m)$, $(n-m)*m$, $(n-m)*(n-m)$, respectively, $V_G$ and $I_G$ refer to the complex voltage and current phasors of synchronous generators, $V_L$ is the complex voltage phasor for the rest of the buses, $m$ and $n$ denote the number of synchronous machines and buses in the system, respectively. The reduced admittance matrix $Y_{red}$ can be calculated by substituting (4) into the equation of generator current in (3) to yield:

$$I_G = Y_{11}V_G + Y_{12}(-Y_{22}^{-1}Y_{21})V_G$$
$$= [Y_{11} - Y_{12}Y_{22}^{-1}Y_{21}]V_G = Y_{red}V_G \quad (5)$$

$$I_G = Y_{red} V_G \quad (6)$$

where

$$Y_{red} = Y_{11} - Y_{12}Y_{22}^{-1}Y_{21} \quad (7)$$

The second MATLAB–function block is to calculate the current commands ($I_{pcmd}$ and $I_{qcmd}$) of the renewable energy plants, which are multiplied by low-pass filters and limiters to produce the currents $I_p$ and $I_q$, respectively. The last MATLAB–function block on the right of Fig. 2b namely "Network" is used to calculate the network admittance–that is, the reduced admittance matrix $Y_{red}$ from (7). Note that $t_{f1}$, $t_{f2}$ in Fig. 2b refer to the times where the disturbance is applied and removed, respectively, and PI_P, PI_I are the gain vectors of the PI controllers, respectively.

## 2.4 Initial Values Calculation

In this study, load flow analysis is carried out using Matpower package [14] to compute the initial values of bus voltage phasors. For the state variables, all the derivative terms in the differential equations (10)–(20) in Appendix II are set to zero and the initial values of these state variables are calculated following the steps described in [15].

## 3. SIMULATION EXAMPLE

### 3.1 Time–Domain Simulation

In this study, the IEEE 68–bus 16–machine dynamic test system is employed to verify the performance of the developed program. There are five areas in this system. Generators G1–9, G10–13 are located in Area 1 and 2, respectively, whereas the other three generators are inside Areas 3–5, respectively. For this research study, the last three generators G14–16 are replaced by three renewable generators modeled using the generic model that could represent wind turbines, solar plants, or an energy storage plant. A three-phase fault is applied at Bus 17 which is connected to the largest generator in the system with a capacity of 3.59 GW. The simulation is completed in a few seconds and the results are shown in Fig. 3.

### 3.2 Modal Analysis

Using the control tools in Simulink, the system is linearized around an operating point and the matrices A, B, C, and D are calculated. Then, the function "*eig*" in Matlab is employed to compute the eigenvalues of the system ($\lambda = \sigma \pm \omega$) and the right-left eigenvectors from these matrices, where $\sigma$ is the real part of the eigenvalue ($\lambda$) and $\omega$ is the imaginary part of $\lambda$. Then, the frequency ($f$ in Hz.) and damping ratio ($\zeta$ in %) for each oscillatory mode is computed from the formulas $\left(f = \frac{\omega}{2\pi}\right)$ and $\left(\zeta = -\frac{\sigma}{\sqrt{\sigma^2 + \omega^2}} * 100\%\right)$, respectively. The system has five lightly-damped modes with damping ratios of less than 10%. These modes with their modal specifications such as eigenvalues, damping ration, and frequency are shown in Fig. 4 with their mode shape plots discussed in the following section.

### 3.3 Mode Shape

From the right eigenvector, one can easily plot the mode shapes of each mode in the system. For instance, one can observe the interaction between the machines in the system corresponding to each mode specifically the lightly-damped modes with a damping ratio of less than 10%. From these plots, we can determine which machine swings against the others and whether the modes are local, inter-area, or other types of modes. Fig. 4 shows the mode shapes of the rotor speeds corresponding to the lightly-damped modes in the system. The lengths of the arrow tell us how much those states contribute in these modes whereas the directions of the arrows provide the displacement angles of the state variables. There are four local modes in the system with a frequency greater than 1.0 Hz. and one inter-area mode with a frequency of less than 1.0. From the inter-area mode shape, the rotor speed states for the machines in the same area are in the same direction and swing against the machines in the other area. For the local modes, we can see some machines in one area swing against some other machines in the same area.

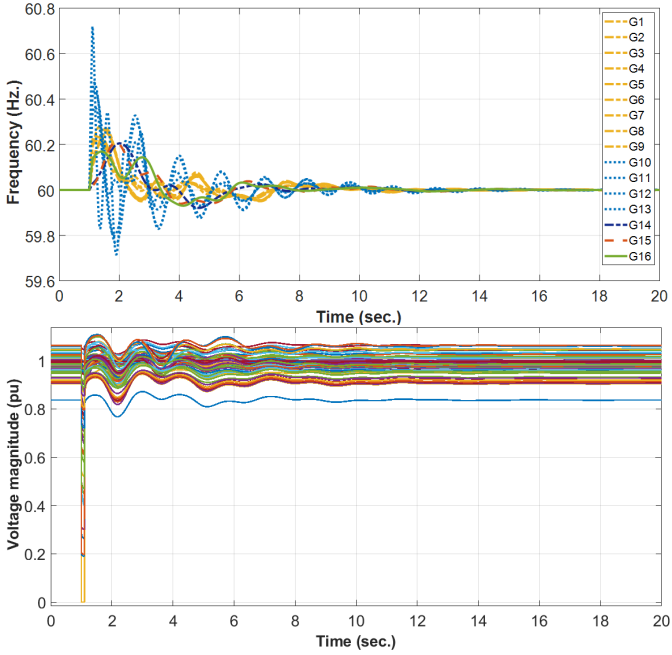

Fig. 3. Frequency (top) and voltage magnitude plots of the system

3.4 Participation Factor Analysis

The relation between each state variable and each mode in the system can be studied through participation factor analysis. Mathematical, participation factor of each state variable can be computed by multiplying the right and left eigenvectors of a specific mode divided by their summation. It can be normalized if divided by the maximum participation factor in the system as follows [16]:

$$p_{ki} = \frac{|v_{ki}||\omega_{ki}|}{\sum_{k=1}^{n}|v_{ki}||\omega_{ki}|} \quad (8)$$

$$p_{ki\_normalized} = \frac{p_{ki}}{max|p_{ki}|} \quad (9)$$

where $v_{ki}$ and $\omega_{ki}$ are the right and left eigenvectors of the system matrix, respectively, and $p_{ki}$ is the participation factor of the $k^{th}$ state variable into the $i^{th}$ mode.

The numerical results of this analysis are stored in an excel sheet and visualized as a map with the state variables on the x-axis and the modes' specifications (damping ratios and frequencies) on the y-axis as shown in Fig. 5a. Since there are 13 synchronous machines in the system and each machine has 11 states (assuming the detailed model of machines equipped with turbine and excitation systems), there are $13*11 = 143$ state variables in the system. However, the most important state variables in the system are the electromechanical states represented by the rotor angles of the machines and their speeds–that is, $\delta$ and $\omega$. Fig. 5b displays this part of the map as highlighted as a red regular in Fig 5a. The values of normalized participation factors are visualized using different colors shown in the bar at the right. Notably, G1 and G10 are the most active machines in the system contributing to the two lowest damping ratios (6.9% and 7.21%). The next mode with a damping ratio of 7.24% is excited by the state variables of G12–13, whereas the mode with the damping ratio of 9.45% is affected mostly by G11. Finally, G13 then G12 contribute largely to the inter-area mode. Owing to space limitation, the numerical results are not shown in the paper but are generated by the program and stores as an excel sheet.

3.5 Modal Controllability, Modal Observability, and Modal Residue

Another important consideration in power system dynamic analysis and controller design is modal controllability ($w_i B$), modal observability ($C v_i$), and modal residue $R_i = C v_i w_i B$, where $R_i$ denotes the modal residue for mode $i$, and $v_i, w_i$ denote the right and left eigenvectors of the system, respectively. These control quantities provide us with information about the inputs\outputs of the linearized system that cannot be obtained through participation factor analysis. It is usually used for two important topics in power systems: optimal location of controller and optimal feedback signal selection.

Figure. 6a–b shows the normalized magnitude and phase angle displacement of modal controllability, observability, and residue of the system. From Fig. 6a, we can find the optimal location of a power system stabilizer to be installed for damping oscillations whereas from Fig. 6b we can determine the amount of phase compensation we need for the lead-lag compensator. Clearly, the best controller locations in the system with the highest residue magnitude are G10, G8, then G1, which demonstrates the results obtained from the participation factor analysis. Note that the transfer function is computed between the rotor speed of G1 and the nominal reference value ($\omega_s$).

3.6 Frequency response Analysis

The developed program can also be employed to plot the frequency response analysis of the system including Bode, Nyquist, and Nichols diagrams as shown in Fig. 7. These plots are used to determine the closed-loop system stability in terms of open-loop gain and phase margins which are required to be positive and large enough. The critical point for the Nyquist and Nichols is (-1, 0). For a closed-loop to be stable, the Nyquist plot of the open-loop transfer function must not make an encirclement around the critical point whereas the Nichols plot needs to be far from this point. The results displayed in Fig. 7 show that the minimum gain and phase margins are 44.9 dB and $119^o$ at frequencies 19.6 and 0.7 rad\s, respectively, and the system is stable. Note that in the pole-zero map, there is no zero-eigenvalue due to abstracting the reference angle–which is corresponding to machine 1 in this study– from all angles [13]. If this reference angle is removed, then a zero eigenvalue will appear in the zero-pole map.

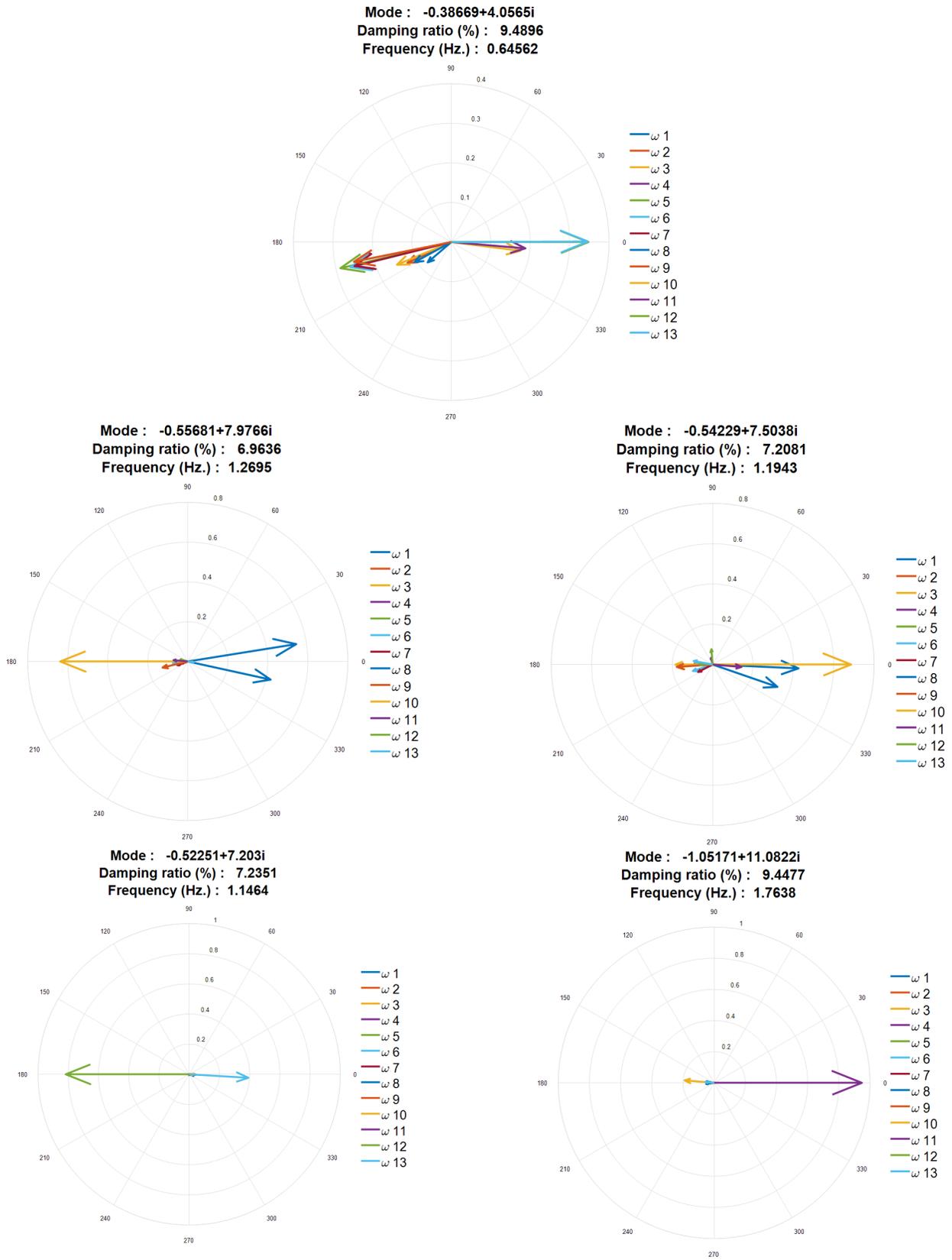

Fig. 4. Mode shape plots of rotor speeds for lightly–damped modes

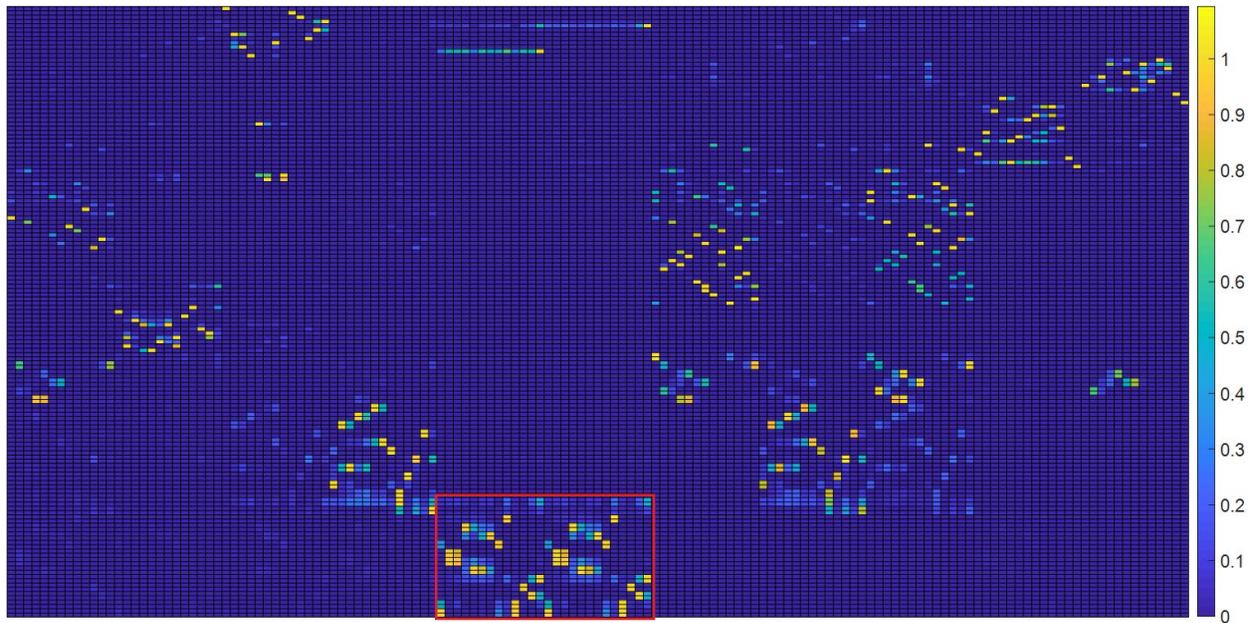
Fig. 5a Participation factor map of the entire system

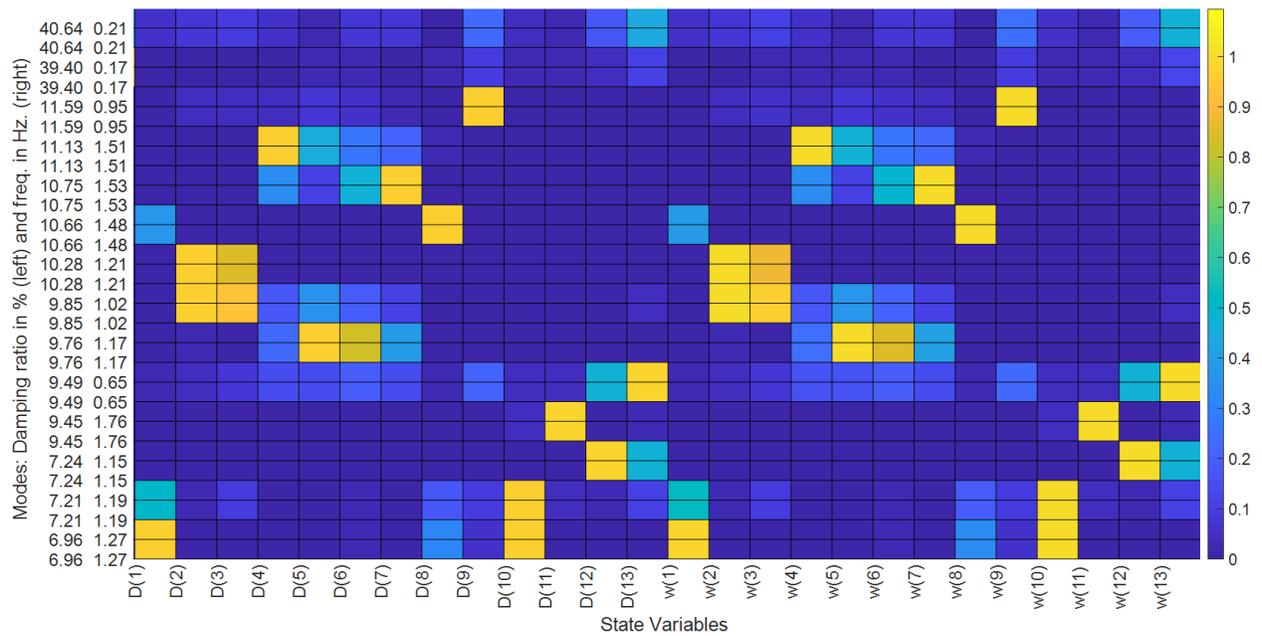
Fig. 5b Magnified part of the map highlighted in red rectangular in Fig. 5a

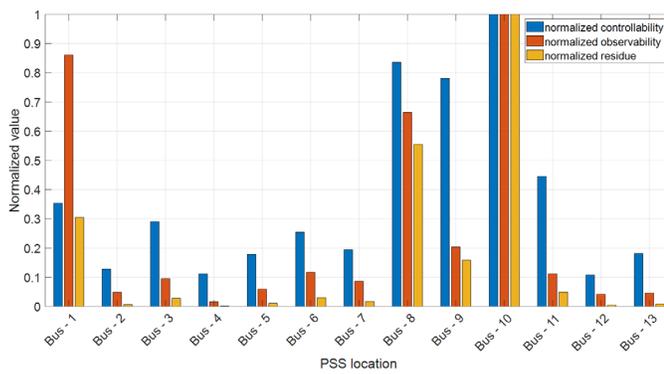
Fig. 6a. Normalize controllability, observability and residue.

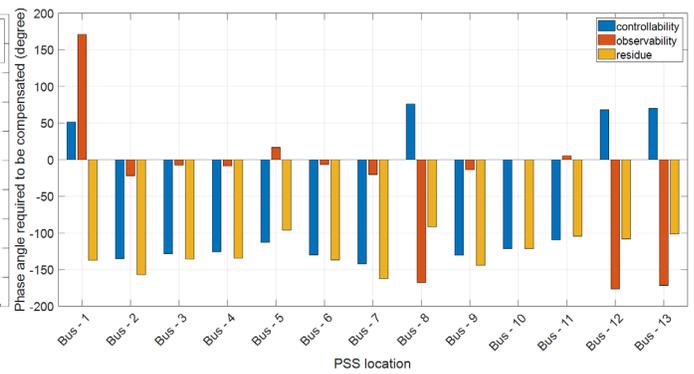
Fig.6b Phase angle displacement

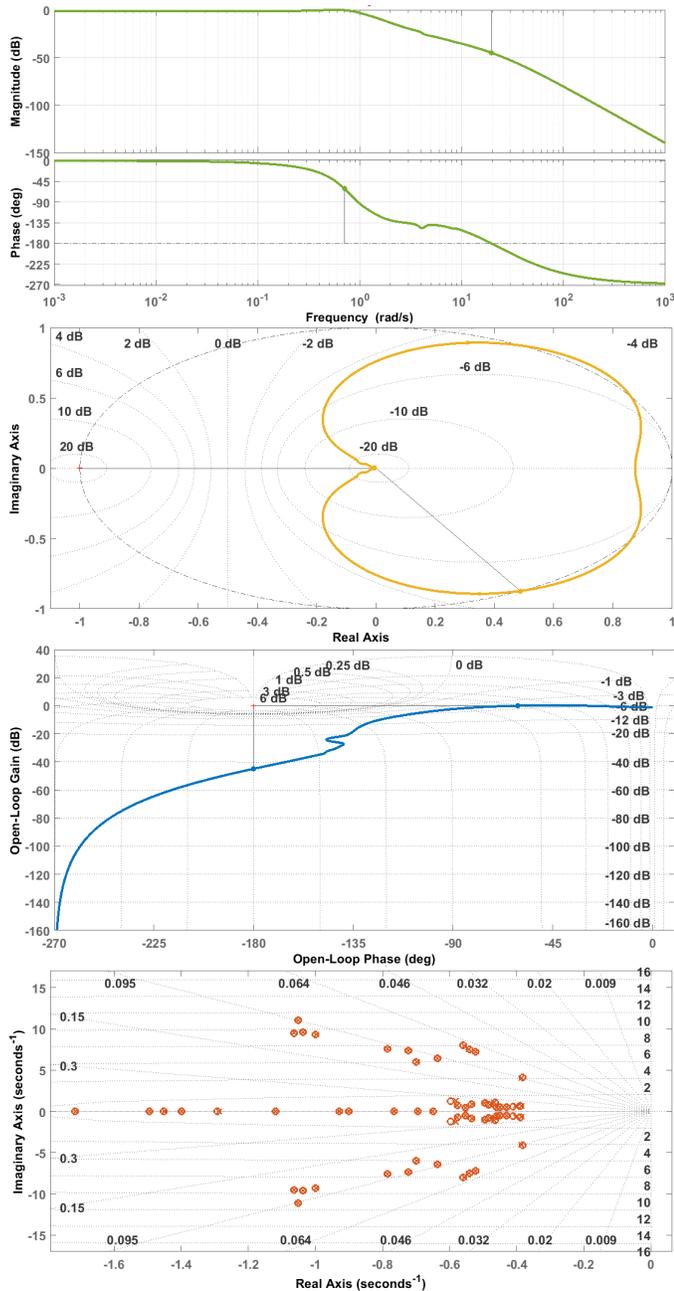

Fig. 7. Bode plot, Nyquist plot, Nichols plot, and pole-zero map of the system

## 4. CONCLUSION

This paper introduced an open-source program developed to model and integrate renewable energy sources into the grid using Simulink. The model can be used for simulating large-scale dynamic modern power systems such as IEEE 68–bus test system employed in this paper. The program can be used for research and educational purposes. Several important topics were covered in this paper using the program including modal analysis, participation factor analysis, optimal location of controllers using modal residue analysis, and frequency response analysis. Comparing to the previous proposed model in Simulink, the proposed model is fast and efficient that takes only a few seconds to simulate the test system with the detailed model of machines and controllers.

## Appendix I

### Abbreviations

| Symbol | Description |
|---|---|
| $R_s$ | Stator resistance in pu |
| $X_d$ | d-axis reactance in pu |
| $X_d'$ | Transient d-axis reactance in pu |
| $X_d''$ | Sub-transient d-axis reactance in pu |
| $X_q$ | q-axis reactance in pu |
| $X_q'$ | Transient q-axis reactance in pu |
| $X_q''$ | Sub-transient q-axis reactance in pu |
| $H$ | Shaft inertia constant in s |
| $w_s$ | Generator synchronous speed in rad per second |
| $T_{do}'$ | d-axis time constant associated with $E_q'$ in second |
| $T_{do}''$ | d-axis time constant associated with $\Psi_{1d}$ in second |
| $T_{qo}'$ | q-axis time constant associated with $E_d'$ in second |
| $T_{qo}''$ | q-axis time constant associated with $\Psi_{2q}$ in second |
| $T_A$ | Amplifier time constant in s |
| $T_{CH}$ | Incremental steam chest time constant in s |
| $T_{SV}$ | Steam valve time constant in s |
| $K_A$ | Amplifier gain |
| $K_E$ | Separate or self-excited constant |
| $E_q'$ | q-axis transient internal voltages in pu |
| $E_d'$ | d-axis transient internal voltages in pu |
| $E$ | Internal voltage in pu |
| $\Psi_{1d}$ | Damper winding 1d flux linkages in pu |
| $\Psi_{2q}$ | Damper winding 2q flux linkages in pu |
| $\delta$ | Rotor angle in rad |
| $w$ | Angular speed of generator in rad per second |
| $\bar{V}_i$ | Complex voltage phasor |
| $V$ | Magnitude of bus voltage in pu |
| $\theta$ | Angle of bus voltage in rad |
| $\bar{I}_{Gi}$ | Generator complex current phasor |
| $I_{Gi}$ | Generator current magnitude in pu |
| $\gamma_i$ | Generator current angle in rad |
| $I_d$ | d-axis current in pu |
| $I_q$ | q-axis current in pu |
| $\alpha_{ik}$ | Angle of admittance $Y_{ik}$ in rad |
| $E_{fd}$ | Field voltage in pu |
| $V_R$ | Exciter input in pu |
| $R_F$ | Rate feedback in pu |
| $T_M$ | Mechanical input torque in pu |
| $P_{SV}$ | Steam valve position in pu |
| $P_C$ | Control power input in pu |
| $R_D$ | Speed regulation quantity in Hz/pu |
| $V_{ref}$ | Reference voltage input in pu |
| $S_E$ | Saturation function |
| $T_{FW}$ | Frictional windage torques |

## Appendix II

### Dynamic Equations of the System

1. Synchronous Generators

$$T_{doi}' \frac{dE_{qi}'}{dt} = -E_{qi}' - (X_{di} - X_{di}')\left[I_{di} - \frac{(X_{di}' - X_{di}'')}{(X_{di}' - X_{ls})^2}(\Psi_{1di} + (X_{di}' - X_{ls})I_{di} - E_{qi}')\right] + E_{fd} \quad (10)$$

$$T_{doi}'' \frac{d\Psi_{1di}}{dt} = -\Psi_{1di} + E_{qi}' - (X_{di}' - X_{ls})I_{di} \quad (11)$$

$$T_{qoi}' \frac{dE_{di}'}{dt} = -E_{di}' + (X_{qi} - X_{qi}')\left[I_{qi} - \frac{(X_{qi}' - X_{qi}'')}{(X_{qi}' - X_{ls})^2}(\Psi_{2qi} + (X_{qi}' - X_{ls})I_{qi} + E_{di}')\right] \quad (12)$$

$$T_{qoi}'' \frac{d\Psi_{2qi}}{dt} = -\Psi_{2qi} - E_{di}' - (X_{qi}' - X_{ls})I_{qi} \quad (13)$$

$$\frac{d\delta_i}{dt} = w_i - w_s \quad (14)$$

$$\frac{2H_i}{w_s}\frac{dw_i}{dt} = T_{Mi} - \frac{X_{di}'' - X_{ls}}{(X_{di}' - X_{ls})}E_{qi}'I_{qi} - \frac{(X_{di}' - X_{di}'')}{(X_{di}' - X_{ls})}\Psi_{1di}I_{qi} - \frac{(X_{qi}'' - X_{ls})}{(X_{qi}' - X_{ls})}E_{di}'I_{di} + \frac{(X_{qi}' - X_{qi}'')}{(X_{qi}' - X_{ls})}\Psi_{2qi}I_{di} - (X_{qi}'' - X_{di}'')I_{di}I_{qi} - T_{FW} \quad (15)$$

2. Excitation systems (IEEE type I)

$$T_{Ei}\frac{dE_{fdi}}{dt} = -\left(K_{Ei} + S_{Ei}(E_{fdi})\right)E_{fdi} + V_{Ri} \quad (16)$$

$$T_{Fi}\frac{dR_{fi}}{dt} = -R_{fi} + \frac{K_{fi}}{T_{fi}}E_{fdi} \quad (17)$$

$$T_{Ai}\frac{dV_{Ri}}{dt} = -V_{Ri} + K_{Ai}R_{fi} - \frac{K_{Ai}K_{fi}}{T_{fi}}E_{fdi} + K_{Ai}(V_{refi} - V_i) \quad (18)$$

3. Turbine systems

$$T_{CHi}\frac{dT_{Mi}}{dt} = -T_{Mi} + P_{SVi} \quad (19)$$

$$T_{SVi}\frac{dP_{SVi}}{dt} = -P_{SVi} + P_{Ci} - \frac{1}{R_{Di}}\left(\frac{w_i}{w_s} - 1\right) \quad (20)$$